# Acquiring IT Solutions through Open Source Software

Mohammad Nabil Almunawar
Faculty of Business, Economic & Policy Studies
Universiti Brunei Darussalam
nabil.almunawar@fbeps.ubd.edu.bn

*Abstract*

Open source software is free software that provides user freedom to use, replicate, modify, and distribute for any purpose. The quality of well-known open source software is very high and they are used by big companies such as IBM, Google and Amazon.com. Recently the number of open source software project growing very fast, which indicates that adoption of open source software is growing although still limited. Businesses should consider open source software as alternative solutions to their business problems or opportunities. An example of a very good open source software for office suite is discussed and compared with the well-known proprietary counterpart.

## 1. Introduction

The is no question regarding the important role of computers in business these days, even micro businesses use computers at least to support their administrative activities such as using word-processing, spreadsheet or any office tool. Larger businesses depend on computers to operate or even to survive.

In general a computer is composed of hardware and software. Hardware technology has improved significantly and the cost to acquire hardware is getting cheaper every year. It is now possible to purchase a powerful machine just for few thousand dollars.

Despite hardware is getting cheaper and more powerful, the cost to acquire software is expensive. Normally software is acquired through licensing. A licence is normally applied for one machine on yearly basis. Many small enterprises cannot afford to buy software licences; however, they need the software to help them running the business. Because of this problem, some small businesses use pirated software. But there is a danger of using pirated software for business since the vendor of the software can sue the enterprise which use pirated copies of its software and demands the enterprise to pay a high price.

An alternative to solve the software acquisition problem is to use Open Source Software (OSS). OSS is free software that provides users freedom to use it for any purpose. In addition the user of OSS has the right to replicate, modify and redistribute copies of the original or modified software freely.

Recently, there are many qualities OSS around. Unfortunately, relatively few businesses adopt OSS. There are few reasons. First, cheap pirated copies of proprietary software are widely available. Second, proprietary software, especially Microsoft operating system and its main application suite have achieved overwhelming market share, which is over 90% among PC users. Third, most new PCs are pre-installed with Microsoft operating system and its office suite. Fourth, there is a conservative bias of many businesses toward proprietary software, especially Microsoft's products. Some other obstacles of adopting OSS are lack of knowledge about OSS, incorrect perception about OSS such OSS is unreliable, insecure and difficult to use, resist to change because of unable to grasp benefits of OSS or simply unnecessary fear and lack of support for users (DiBona, Ockman, Stone, Behlendorf, Bradner, Hamerly, et al., 1999)

Despite some hindrance factors mentioned above, some OSS start gaining wide acceptance, to name few are Linux, Apache, Sendmail, PHP, MySQL and Firefox. Linux is a very reliable and powerful operating system that can run on less powerful PCs. A well-known website, Google, has deployed thousands Linux servers. Apache is a number one web server since 1996. In recent survey (January 2009) conducted by E-Soft Security Space, Apache still dominates web server over the Internet with 72.45% share, followed by its main competitor, Microsoft Internet Information Server (IIS) with 18.05% share[1]. Sendmail is the leader in email server, PHP is the most popular server-side scripting language, MySQL is a powerful database management system that is currently growing fast, and Firefox is light compare to Microsoft Internet Explorer, but it is a powerful browser start gaining popularity for surfing the Internet.

Businesses should consider OSS as alternative solutions to their business problems. What businesses need is study their requirements and to find out whether OSS can fulfill them or not. This is important since several OSS have sufficient high quality and have nudged out their proprietary counterparts. It is also important to note that several well-known players in the Web such as Google, Amazon and Salesforce.com take advantage of reliability and low cost of OSS to build a platform that can create more businesses for them. Once one decides to use OSS one needs to prepare several important steps, get the software and install it properly, user training, and managing change. Finding support is another important thing need to be taken care.

## 2. Open Source Software

The history of open source can be traced during early development of large scale commercial computer in 1950s and 1960s where software was free and came with source code. The source code can be modified and recompiled to improve the software. At this time software was not considered as a revenue generator, it was just a necessary ingredient to make hardware work and to provide useful functionalities or solutions so that people are encouraged to buy expensive hardware (mainframes).

---

[1] http://www.securityspace.com/s_survey/data/200812/index.html

The term Free Software (FS) has been introduced by Richard Stallman (Stallman & Lessig, 2002). He is the one who established GNU Project and Free Software Foundation. GNU general Public Licence (GNU GPL) has been introduced. This licence guarantees that software under this licence is free. The term open source software was introduced in 1998 in a brainstorming session among some notable persons in this area in Paulo Alto, California[2]. The term *open source software* used instead of *free software* to embrace business world (Perens, 1999)

The freedom of using and modifying software came to an end by late 1960s when IBM started charging separately for software. By mid 1970s much of software distributed was no longer free and source code was inaccessible. Another example of switching from open to close is UNIX, an operating system for mainframes developed by AT &T. It was initially free and the source code was available. However, in 1980s AT&T started commercialising UNIX and asked users, even from universities, to sign non disclosure agreements.

Since it commercialisation mentioned above, software business has become a giant business that generates multi-billion dollars every year. Bill Gates, a number one person of Microsoft has become the number one on the "Forbes 400" list from 1999 to 2007. Licensing software is the major revenue source of software companies. So, why there exists groups of people or movements that strongly oppose the idea of making money from software licence and then subsequently offer free software? There must be strong motives or belief that underpins such groups or movements.

Free software movement championed by Stallman was the reaction to the policy of non-disclosure of codes enforced by commercial motives from software companies. As a matter of fact programmers were unable to fix bugs created by programmes that they supervised, which in turn created frustrations for them. However, there was a deeper concern regarding creativity and control. Programmers, especially hackers, used to study in detail source code of software. Understanding of source code will eventually allow them to modify and improve the software. With the non-disclosure policy of software, hackers are unable to study the source codes.

Non-disclosure policy also means control that is the complete control of software companies to their users. In this regards, users basically do not have any freedom to the software they licensed. If user wants to have any changes of software, even a small one, they have to request the owner of the software, and normally this change will cost users a lot of money. Users even cannot help their fellows to solve their problems by copying the software. Stallman said (Stallman, 1999; Stallman & Lessig, 2002) "If you share with your neighbour, you are a pirate. If you want any changes, beg us to make them." He accused proprietary software companies are anti-social since they don't allow user to share or change software. Sharing, especially information-sharing is one of the core hackers belief. Mackenzie (Mackenzie, 2001) in his books review quotes this core belief from Raymod (1996), who said that "information-sharing is a powerful positive good, and it is an ethical duty of hackers to share their expertise by writing free software and facilitating access to information and to computing resources wherever possible."

---

2  Http://www.opensource.org/docs/history.php

The idea of free software from Stallman is basically a strong reaction to non-disclosure policy of proprietary software companies. According to Stallman "Free" in the free software refers to freedom, not refer to price. He defines a program is free software if

- *a user has the freedom to run the program, for any purpose.*
- *a user has the freedom to modify the program to suit ones needs. (To make this freedom effective in practice, a user must have access to the source code, since making changes in a program without having the source code is exceedingly difficult.)*
- *a user has the freedom to redistribute copies, either gratis or for a fee.*
- *A user has the freedom to distribute modified versions of the program, so that the community can benefit from ones improvements.*

Further, Stallman (Stallman & Lessig, 2002) introduced the term *copyleft* contrasting the term *copyright* of proprietary software. *Copyleft* is a distribution concept to make sure that software under this term is always free. Note that a modified version of *copyleft* software is under *copyleft* as well, which means it must be kept as free software.

The open source movement shares the basic principles of Stallman's free software movement. As Stallman (2002) said "free software" and "open source" describe the same category of software, more or less, but say different things about the software, and about values. He further said that "The Free Software Movement and Open Source Movement are two political parties in the same community."

Open Source Movement has taken different path from Free Software Movement to promote the idea of free software. Notable persons in Open Source Movement such as Eric Raymond, Bruce Perens, Tim O'Reilly and others concerned about an anti-business message from Free Software Foundation (DiBona et al., 1999) They were worry that idea of free software and quality software produced would be sidelined. They agreed that to promote the idea, they should embrace business world. The term "free software" is not business friendly, so they agreed to introduced the term *open source software.* The new term attracted a lot of support from hacker culture (Raymond, 1998) Subsequently, supports also came from software companies such as Netscape, Sun Microsystems and IBM. The open source definition is taken from Debian Free Software Guidelines[3]. A software is considered as an open source

---

[3] http://www.debian.org/social_contract.html#guidelines

software if it complies 10 requirements stated in the definition. The 10 requirements are (taken from http://www.opensource.org/docs/definition.php):

1. ***Free Redistribution.*** *The license shall not restrict any party from selling or giving away the software as a component of an aggregate software distribution containing programs from several different sources. The license shall not require a royalty or other fee for such sale.*

2. ***Source Code.*** *The program must include source code, and must allow distribution in source code as well as compiled form. Where some form of a product is not distributed with source code, there must be a well-publicized means of obtaining the source code for no more than a reasonable reproduction cost–preferably, downloading via the Internet without charge. The source code must be the preferred form in which a programmer would modify the program. Deliberately obfuscated source code is not allowed. Intermediate forms such as the output of a preprocessor or translator are not allowed.*

3. ***Derived Works.*** *The license must allow modifications and derived works, and must allow them to be distributed under the same terms as the license of the original software.*

4. ***Integrity of The Author's Source Code.*** *The license may restrict source-code from being distributed in modified form only if the license allows the distribution of "patch files" with the source code for the purpose of modifying the program at build time. The license must explicitly permit distribution of software built from modified source code. The license may require derived works to carry a different name or version number from the original software.*

5. ***No Discrimination Against Persons or Groups.*** *The license must not discriminate against any person or group of persons.*

6. ***No Discrimination Against Fields of Endeavour.*** *The license must not restrict anyone from making use of the program in a specific field of endeavour. For example, it may not restrict the program from being used in a business, or from being used for genetic research.*

7. ***Distribution of License.*** *The rights attached to the program must apply to all to whom the program is redistributed without the need for execution of an additional license by those parties.*

8. ***License Must Not Be Specific to a Product.*** *The rights attached to the program must not depend on the program's being part of a particular software distribution. If the program is extracted from that distribution and used or distributed within the terms of the program's license, all parties to whom the program is redistributed should have the same rights as those that are granted in conjunction with the original software distribution.*

9. ***License Must Not Restrict Other Software.*** *The license must not place restrictions on other software that is distributed along with the licensed* *software. For example, the license must not insist that all other programs distributed on the same medium must be open-source software.*

10. ***License Must Be Technology-Neutral.*** *No provision of the license may be predicated on any individual technology or style of interface*.

Basic principles of OSS/FS movements have been captured in the definitions of free software and open source software above. We would like to highlight some motives that underpin programmes participating in OSS/FS projects. Several studies on motivation for participating in OSS/FS projects (Hars & Ou, 2002; Hertel, Niedner, & Herrmann, 2003; Lakhani & Wolf, 2003) classify motivations into two types: intrinsic and extrinsic motivations. Intrinsic motivation includes altruism, personal challenges, having fun, enhancing skill, obligation to contribute back to community. Extrinsic motivations can be grouped into four main theme, 1) own use, 2) signalling effects, such as recognition and reputation, 3) intellectual challenges, and 4) ideological source of motivation (Dahlander & McKelvey, 2005)

A survey on hackers participation on OSS projects ported in SourceForge.net conducted by Boston Consulting Group (Lakhani, B. Wolf, Bates, & DiBona, 2002) found that the main motivations are *intellectually stimulating* and *improve skills*. The Overall motivations of hackers can be seen in Figure 1.

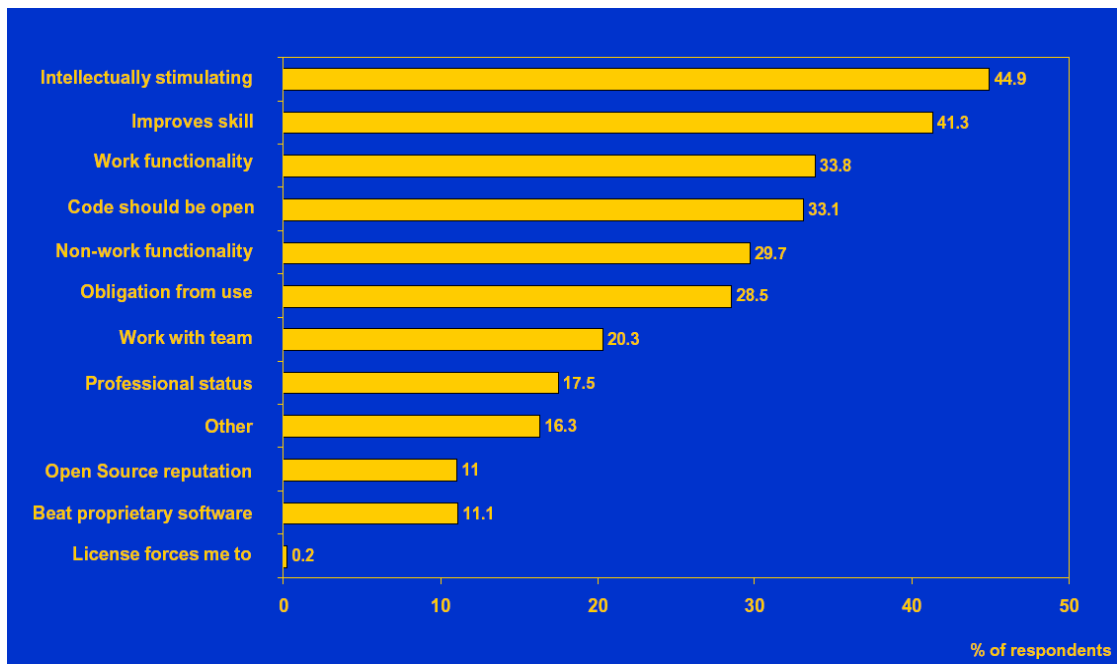

Figure 1: Overall motivations of hackers (courtesy of Lakhani at al., 2002)

## 3. Quality of OSS/FS

The first thing that relate to the quality of a software is reliability. Unreliable software implies that that software contains many bugs, which definitely very disturbing or even dangerous.

In a normal daily life free or low cost products or services highly associate with their low quality. However, this cannot be applied to OSS/FS. Although OSS/FS is free, the quality of a stable version (production release) of OSS/FS is normally very high. It may surpass its proprietary counterparts. In fact well-known OSS/FS such as Linux, Apache, MySQL and OpenOffice are high quality and very reliable software. It is interesting to note that Forrester Research has recommended that "firms should consider open source options for mission-critical applications." (Golden, 2005). This indicates that the quality of open source software is sufficiently high so that even for mission-critical applications it should be considered.

There several research reporting reliability of OSS/FS as compare to proprietary software. Miller, Cjin, Maganty, Murthy, Natarajan, & Steidl (1995) in their work examining the reliability of UNIX utilities and services conducted reliability tests over 80 utility programs on nine different UNIX platforms, seven of these were commercial platform and the rest were GNU/Linux and GNU tools. The testing method was quite simple, by inputting random input streams generated by a program called the *fuzz generator* to the utilities tested. The result of tests were very interesting. OSS/FSS utilities are more reliable than proprietary utilities. The failure rate (because of bugs) of proprietary utilities ranged from 15 to 43% (the average failure is 23%) while the failure rate of Linux was 9% and the failure rate of GNU utilities was only 6%.

The same method of test was conducted for Windows NT and Windows 2000 by Forrester and Miller (2000). They tested 30 GUI-based applications by inputting them with streams of valid keyboard, mouse events and streams of random Win32 messages generated by the *fuzz generator*. The results were 21% Windows NT applications crashed and 24% of them hung. So, the average failure of cashed and hung was 22.5%. However when those applications were fed with completely random input streams consisting of random Win32 messages, the failure rates rose up to 100%.

Recently Miller, Cooksey, & Moore (2007) conducted a similar reliability testing on GUI applications running on MacOS X using the same method as discussed above. Failure rate of command line utilities was 7% . However, failure rate of GUI-based utilities is surprisingly high, 73%. Note that MacOS X is a major step for Apple to switch to UNIX-based operating system.

There are few other studies summarized by Wheeler (2007) shown that OSS/FS are higher quality than their proprietary counterparts. For example, OSS/FS networking reliability is outperform many proprietary counterparts. Shankland (2003) reported that the quality of a key networking component of GNU/Linux is superior compare to many proprietary operating systems. A similar study comparing code quality of MySQL, a popular OSS/FSS database management systems to 200 proprietary counterparts found that MySQL has defect density only 0.09 (0.09 defects/1000 lines of code) while the proprietary counterparts have average defect density 0.57 (Wheeler, 2007). This shows that OSS/FS is of high quality.

An interesting study on comparing maintainability of code OSS/FS to proprietary software was conducted by Samoladas, Stamelos, Angelis, & Oikonomou (2004). The study found that the maintainability both OSS/FS and proprietary software deteriorate overtime. However, maintainability of OSS/FS does significantly better than its proprietary counterpart. Regarding code quality the study concludes that mentioned that the code quality of OSS/FS comparable to or sometime better than the quality of proprietary software code implementing the same functionality.

There are few reasons as to why the quality OSS/FS (production release) are high and more reliable that proprietary software although OSS/FS lack of formal testing support and does not follow software engineering standards. Many OSS/FS programmers are highly skilled and motivated programmers, Figure 1 support this. (Miller et al., 1995) argue that OSS/FS has a personal touch between authors and users which improve communications between them. Bugs found by users can be reported immediately and they can get immediate response from the authors. Proprietary software, which normally produced by big companies does not have such mechanism. Big company tend to slow in responding of users bugs report and the user never see or hear about the resolution.

In addition, the two principles of OSS/FS "release early and release often" and "given enough eyeballs all bugs are shallow" are the main determination for quality of OSS/FS. Life cycle of OSS/FS are shorter than proprietary software and bugs can be fixed on each cycle. By involving many people in a transparent development process, bugs or defects can be easily and quickly spotted.

**4. Considering OSS/FS for business solution**

As mentioned above, although there is no cost for OSS/FS licences the quality of stable versions of well-known OSS/FS are very high. Furthermore the number and variety of OSS/FS projects are growing fast. For example, there are more than 160 thousands active OSS/FS projects registered in SourceForge.net alone in October 2007 (http://SourceForge.net). Therefore, overlooking OSS/FS in making a decision to acquire software is ignorant.

In general there are few steps in making decision to acquire software for business solutions. The first step is to define your business needs. The outputs of this step are features or functionalities needed to solve the business problems.

The second step is to identify alternative solutions given the functionalities required. In this step you should include OSS/FS. You can search alternative solutions through a well-known search engine like Google or AltaVista. For OSS/FS solutions you can search them on OSS/FS projects repositories such as SourceForge.net, freshmet.net,

open source list (http://www.opensourcelist.org), savanah (http://savanah.gnu.org) or free software directory (http://directory.fsf.org). A dedicated OSS/FS run on windows environment can be found in OSSWin projects (http://osswin.sourceforge.net/). To add knowledge on selected alternative, you can read existing reviews on those alternatives. Reviews may also give you ideas additional things need to be considered. Definitely reviews will help you to compare the alternative solutions.

The third step to compare alternative solutions. This step help you to narrow down alternatives found in the previous step so that you can perform an in-depth analysis on a reasonable number of alternative to make final decision. You should compare alternative solutions using criteria which include, functionality, performance, reliability, cost, popularity, support, scalability, flexibility, security, connectivity, maintenance, documentation, hardware requirement, and other factors that you think important to include such licence issues.

The fourth step is to do in-depth analysis short-listed candidates taken from the previous step. This may involve testing the short-listed candidates on critical factors to find strengths and weaknesses of each candidate. You should make sure that the output of this step is the best alternative that can fulfil your business needs, the cheapest and provide sufficient support.

**5. An open source alternative solution for an office suite**

Office suites are main applications used in contemporary work environment. Almost everyone working with computers needs an office suite. An office suite is normally composed of a word processor, a spreadsheet, a presentation tool and a database management system. There are several proprietary as well as open source office suites around. Examples of proprietary office suites are Microsoft Office, WordPerfect Office, StarOffice and Lotus SmartSuite and examples of open source office suites are OpenOffice, Gnome Office, KOffice, and NeoOffice.

Microsoft Office (MS Office) is currently a dominant office suite. It seems there is no serious threat of MS Office leadership on the office suite market at the moment. The popularity of MS Office and the effortlessness to get the unlicensed copies of the suite has made MS Office the prime choice for businesses. However, many cannot afford to purchase over-priced software licences, for examples: non-profit organisations, small businesses and students and because of this, there are many pirated versions of MS Office used around.

Among open source office suites listed above, OpenOffice is the best alternative open source suite. Its interface is similar to MS Office. In addition, it has the ability to read MS Office native document format. In other words, OpenOffice is compatible with MS Office. The following is a brief comparison between MS Office 2003 and OpenOffice 2.3. Note that this comparison is based on a simple testing of MS Office 2003 and OpenOffice 2.3, documentations of MS Office 2003 and OpenOffice 2.3 as well as some information from web based sources, including detail review, comparison and compatibility tests from various PC Magazines such as PC Pro, BYTE, Idealware (www.idealware.com), ZDNet and ComputerWorld.

MS Office 2003 contains five main programs: *Word*, a word processor; *Excel*, a spreadsheet program, *Outlook*, a personal information manager and email client; *PowerPoint*, a popular presentation program; *Access*, a database management system. The suite may include *InfoPath*, a tool for designing XML-based forms; *FrontPage*, a Web design/authoring tool; *Publisher*, software to create newsletters, leaflets, flyers, business cards, etc.; some other tools such as *Visio, Picture Manager* and *Clip Organizer* may also be included. MS Office 2003 has many useful applications, however the most frequently used programs are among the four main ones: *Word, PowerPoint, Outlook* and *Excel*.

Open Office 2.3 contains four main programs: *Writer*, a word processor similar to *Word*; *Calc*, a spreadsheet similar to *Excel*; *Impress*, a presentation program similar to *PowerPoint*; *Base*, a database management system similar to MS Access. There is no comparable program in Open Office 2.0 to *Outlook,* however Mozilla Thunderbird (email client) and Mozilla Sunbird (calender application) are available as free downloads. Some other tools in the suite are *Draw*, a vector graphic editor and *Math*, a mathematical formulae editor. Some other tools include in the version 2.3 are business cards editor, html editor, label editor and XML form editor.

*Cost*
MS Office 2003 will cost you around US$ 500 per licence. Upgrading to the latest version, MS Office 2007 is basically buying a new licence, while OpenOffice 2.3 is free. OpenOffice use LGPL (Lesser General Public Licence) that allows copying, distributing and modifying the software freely. The modified or upgraded version of LGPL is free to be copied, distributed or modified as well. Therefore the upgrade version of OpenOffice is always guaranteed to be free as well, so users do not need to worry about being charged for the next version of the software.

*Performance and System Requirements*
Based on several performance tests, MS Office 2003 is faster and use less RAM than OpenOffice 2.3. But the performance has been improved in OpenOffice 2.3. Running OpenOffice using current PC is not an issue. All programs in OpenOffice can be run with acceptable performance. I tested all programs in OpenOffice 2.3 using a low-end laptop, Acer Aspire 3620 (Intel Celeron M 1.5 MHz, with 256 MB RAM), they run pretty well.

MS Office 2003 professional edition needs higher system requirement. It needs a Pentium III with at least a 233-MHz processor and a minimum 128 MB RAM if it is installed without business contact manager for Outlook or at least 450 MHz and a minimum 400 MB RAM if it is installed with it. It needs Windows 2000 with Service Pack 3 installed, Windows XP or later. OpenOffice 2.3 can run on Windows '98 or higher with a minimum of 128 MB RAM on a Pentium compatible machine. So, OpenOffice can run on less powerful computer.

*Platform, Interoperability and Portability*
MS Office, including MS Office 2003, is designed to run under Windows, however it can also be run on Apple machines, but unfortunately, it is not designed to run on multi-platform environments. OpenOffice 2.3 can be run on various operating systems (multi-platform), including Windows, Mac OS X, GNU/Linux, Sun Solaris (under X11) and FreeBSD. MS Office 2003 uses Microsoft's native document format while OpenOffice uses open document format (ODF) but can import from and export to Microsoft's native document formats. ODF is an XML-based document format dedicated for office applications; it is an international standard approved by International Standard Organization (ISO). An ODF document can be moved across platforms, given the current popularity of flash-drives, it will very useful if a document can be read by any word processor from different platforms without any conversion. This is the main strength of OpenOffice 2.3 over MS Office 2003. MS Office 2007 has entirely different format from previous version. Microsoft has adopted the XML-based format for MS office 2007.

*User Friendliness, Compatibility and Functionality*
Both OpenOffice 2.3 applications and MS Office 2003 applications are very user friendly. OpenOffice 2.3 applications user interface is very similar to the corresponding MS Office 2003 applications user interface. If you are used to working with MS Office applications such as *Word* and *Excel*, you will find that they are very similar to *Writer* and *Calc*. This implies that you do not need to learn much to be able to operate applications in OpenOffice 2.3.

As previously mentioned OpenOffice 2.3 can read from and write to any corresponding MS Office 2003 document, but not the other way around. In fact according to a review carried out by PC Pro on OpenOffice 2.0, *Writer* can read very complex documents created by *Word* seamlessly and *Calc* can read a complex document created by *Excel* smoothly. This means that OpenOffice 2.0, hence also OpenOffice 2.3, is highly compatible with MS Office 2003, hence there is no need to worry about the capability of OpenOffice 2.3 to read existing MS Office documents. However, OpenOffice does not support MS Office macros, this means *Word* macros cannot be run in *Writer* (which may be beneficial when considering the proliferation of macro viruses).

Now, we will look at the key functionalities of the main applications in both suits. We start with *Word* and *Writer*. This article was written using both *Word* and *Writer* changing application several times while observing key functions in both applications. Indeed they are very similar. *Writer* provides useful function to save a document to a PDF format. This means you can generate PDF file from OpenOffice without the help from additional software. *Word* 2003 does not have this function (*Word* 2007 does, though this functionality has been removed from European versions).

Another useful tool in *Writer* is the wizard that guides you to write letters, facsimiles, agendas and many others. Not only it is useful, but it also a very nice tool. Note that *Word* 2003 provides only a wizard to write a letter. Another useful tool in *Writer* is email. You can immediately send the document that you write by clicking the email button. You can even choose to send your document as a *Word*, ODF or PDF document via the file menu. Amazing!

*Writer* includes a digital signature facility. You can digitally sign your document. It is an advanced feature which is not supported by *Word* 2003. Digital signature can be a very useful tool since it helps to preserve the integrity of a document.

*Word* 2003 has spelling and grammar checking facilities, while *Writer* has only a spell checker. This shortcoming can be overcome by an open source/free plug-in tool called *Language Tool*[4]. There are some advanced features that *Word* has but *Writer* doesn't, such as *online collaboration*, *smart tags* and *research pane*. But all these features are rarely used anyway, so, in general *Writer* and *Word* have comparable features. In fact *Write* has several useful features which *Word* 2003 does not have, as discussed above.

Now let us to observe *Excel* and *Calc*. The *Calc* user interface is highly intuitive and has a familiar look and feel, very similar to *Excel's*. *Calc* has many of the *Excel* features and it can read from and write to *Excel's* document format. Furthermore *Calc* has similar useful tools that *Writer* has, such as export document to PDF format, email document in various formats and digital signature facility. So, if you are familiar with *Excel*, you can immediately work with *Calc*, you won't miss anything. According to several compatibility tests between *Calc* and *Excel*, most of the time *Calc* can read *Excel* documents, however it does have some problems in importing graphs from *Excel*.

What about *Impress* and *PowerPoint*? We tested *Impress* by reading several *PowerPoint* files, it reads smoothly all the file without missing anything. It plays files perfectly when running slide shows. *Impress* has a very familiar interface and features which are similar to *PowerPoint*. It also has additional tools as available in *Writer* and *Calc* such as PDF conversion, email and digital signatures. However, *Impress* does not have many presentation templates and ready-made backgrounds like *PowerPoint;* It provides only two presentation backgrounds, but of course you can create or design your own backgrounds or download templates from the Internet.

*Base* is a database management system (DBMS) like *Access*. *Base* is quite similar to *Access*, but *Access* is more mature and has many more features compare to *Base*, also creating commands in Base's forms is rather difficult to do. As a new comer *Base* is a good database application. Unlike the word processor, spreadsheet and presentation tools, you do need a prerequisite knowledge of database design to work with a DBMS. A serious database programmer will work with more powerful and multi-user database management systems such as *DB2, Oracle* or *MySQL* to create more useful applications rather than using a DBMS available in an office suite.

---

4  http://www.danielnaber.de/languagetool/

*User Support*

The first user support is a useful documentation, such as a searchable help file. OpenOffice 2.3 provides you with a comprehensive searchable help file. You can consult problems found while using OpenOffice 2.3 with the help facility in the *Help* menu. Useful documentations are available online from the OpenOffice documentation site (http://documentation.openoffice.org). You can join a related mailing-list and pose your questions there if you have problems.

Furthermore, there are many online tutorials available on the Internet to help you master OpenOffice 2.0. One notable example are the tutorials provided in www.tutorialsforopenoffice.org. This site provides comprehensive tutorials for *Writer, Calc, Impress* and *Draw*. You can download all the tutorial materials in PDF or ODT (ODT extension file for *Writer*) for free.

**6. Conclusion**

OSS is free software that provides users freedom not only to use it for any purpose but also to replicate, modify and redistribute copies of the original or modified software freely.  Although OSS is free, but it does not necessarily sacrifice the quality. The quality of a stable version of well-known OSS is very high and some of them are considered exceed the quality of their proprietary counterparts.

Adoption of OSS in businesses are growing, however many businesses are not aware that OSS can be a relatively cheap solutions for their problems. Businesses need to consider OSS seriously when deciding to acquire IT solutions.

There are so many useful OSS and the number are growing fast. Although the majority of OSS software initially developed on UNIX-like platform, the number of OSS run on multi-platform, including on Microsoft's operating system are growing.

An example of a good OSS software to consider for productivity improvements is OpenOffice. OpenOffice is an office suite similar to MS office. For example, OpenOffice 2.3 user interface is very similar to MS Office 2003's. Almost all features of MS Office 2003 are available in OpenOffice 2.3. It can read MS Office documents. Moreover it has some useful features which are not in MS Office, such as the ability to save documents in the PDF format. Any businesses that are considering to upgrade their office suites should consider OpenOffice.